\documentclass[]{article}

\usepackage[numbers, sort&compress]{natbib}
\usepackage[pdftex]{graphicx}
\usepackage[cmex10]{amsmath}
\usepackage[caption=false,font=footnotesize]{subfig}
\usepackage[usenames,dvipsnames]{color}
\usepackage{url}
\usepackage{fancyvrb}
\input{defs.tex}

\begin{document}
%
\title{Promoting scientific thinking with robots}

\author{Juan Pablo Carbajal,Dorit Assaf, Emanuel Benker}
\maketitle

\begin{abstract}
This article describes an exemplary robot exercise which was conducted in a class for mechatronics students. The goal of this exercise was to engage students in scientific thinking and reasoning, activities which do not always play an important role in their curriculum. The robotic platform presented here is simple in its construction and is customizable to the needs of the teacher. Therefore, it can be used for exercises in many different fields of science, not necessarily related to robotics. Here we present a situation where the robot is used like an alien creature from which we want to understand its behavior, resembling an ethological research activity. This robot exercise is suited for a wide range of courses, from general introduction to science, to hardware oriented lectures.
\end{abstract}

\section{The Braitenberg vehicle exercise}

A simple self-made robotic platform built by the authors was used for the activity. The robot had two wheels, each one actuated by a DC motor. Two light sensors~\cite{LEds} were attached to the robot. The robot was controlled by a simple on-board program that defined a relation between inputs coming from the sensors and output signals sent to each motor.

\begin{figure}[htpb]
\begin{centering}
   \includegraphics[width=0.3\textwidth]{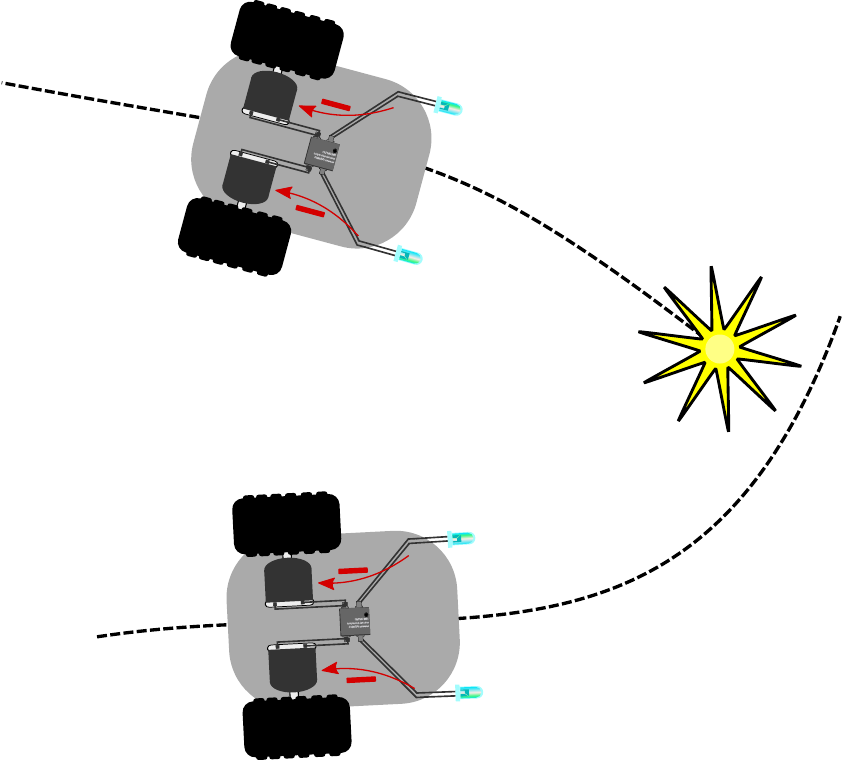}
   \caption{Braitenberg vehicle 3, the light lover. Each sensor reduces the speed of the motor on its side proportionally to the measured light intensity. The figure shows the qualitative behavior of the robot: it moves towards the light and tends to stop close to it.}
   \label{fig:LightLover}
\end{centering}    
\end{figure}

We provided the robot with the behavior of Valentino Braitenberg's vehicle number 3~\cite{Braiten}. The light sensors of the robot commanded the rotational speed of the two motors. The connection was inhibitory, meaning that when the sensor measured light, the speed of the motor connected to it was reduced proportionally to the sensor's output. This sensor-motor configuration generates a light following behavior (Figure \ref{fig:LightLover}). More details about the robot, the control program and how to reproduce this exercise are explained in later sections. 

Next we describe how we used the robot to engage students in scientific thinking. This exercise was part of a class on modeling mechatronics systems that took place at the Baden-Wuerttemberg Cooperative State University Loerrach, Germany. The students were mainly 3rd year bachelor students. The objective of the activity was to let students find out the sensor-motor relationship by means of hands-on experimentation and free exploration. The students had to create a hypothesis about the controller implemented in the robot and later verify it through experiments.

\paragraph{Introducing the robot.} The activity started with the presentation of the robot and a demonstration of its behavior when a light was placed in front of it. The robot moved by default in a straight line, and when it passed close to the light it turned towards it. The robot was even able to track the light (this depends on the sensor gain and motor speed, therefore it requires calibration prior to  demonstration). This light loving behavior, though simple, always captivates the audience as well as the teachers.

\paragraph{The assignment.} After several playful tests with the light, the students were asked to give explanations, in the simplest possible way, about the controller implemented in the robot such that it shows this behavior. Additionally, they were asked to propose an experiment that tests their explanation. In other words, they were asked to develop a model of the internal works of the robot and to produce a hypothesis verifiable through experimentation. The robot allowed us to create a complete and interesting research situation. At this point, to avoid diverging explanations, we suggested to the students to focus on the role of the sensors.

\paragraph{Hands on.} The production of models and tests was done in small groups (3-4 people) and we let the students form the groups by themselves. During this phase, we visited each group and discussed their ideas to assure the experiments will help deciding whether a given model should be discarded or not. It is important to remark that we did not correct the models, since any model is just an approximation. Thus, we just suggested changes in the model to simplify the verification process. After several minutes of group discussion, the groups presented their models, the experiment to be conducted on the robot and what they expected to observe. Since the number of available robots was enough, the students were able to perform their experiments. Otherwise the teacher could select a few experiments and try them in the robot.

\paragraph{The closure.} The conclusion of the activity is left to the criteria of the teacher. In our case, due to the lack of time, we explained the controller and introduced Braitenberg's ideas. In other circumstances, we would have requested the students to produce a short report of the experience and postpone the explanation to the next class.

\section{Robot hardware}

\begin{figure*}
\begin{centering}
	\includegraphics[width=0.6\textwidth]{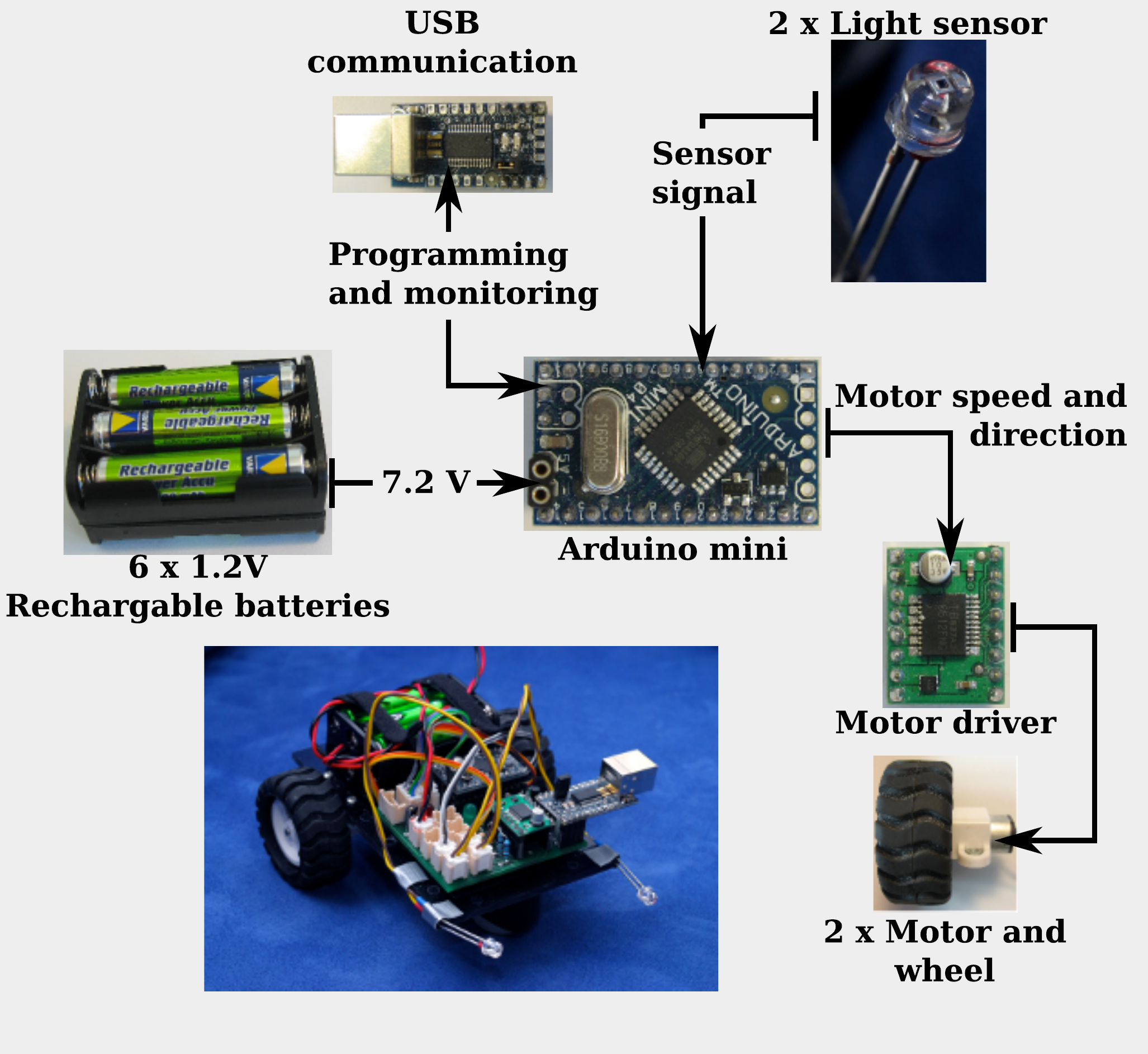}
	\caption{The robot and its components. Six rechargeable batteries are used for power supply. An USB communication unit is used for programming and monitoring the control board. Two light sensors provide sensory input to the control board which controls the two DC motors and wheels through the motor driver component.}
	\label{fig:RobotHardware}
\end{centering}	
\end{figure*}

The custom robotic platform is shown in Figure~\ref{fig:RobotHardware}. Next, we describe the hardware that is needed to reproduce the robot exercise just described. 

As mentioned above, the robot has two motors that can rotate individually at different speeds. Light sensors are placed at the right and left front side of the robot. These sensors can detect a light source within a range of about 10 cm and were previously calibrated by the students by measuring the output voltage as a function of the distance to a light source. The robot control program was implemented such that each light sensor is {\it connected} (via the controller) to one motor and influences its speed directly. Whenever a light sensor measures light the speed of the motor is reduced proportionally to the sensor's measurement. The less light a sensor detects, the faster the motor rotates and vice versa.

A commercial Arduino control board (\url{http://www.arduino.cc}) was used to control the robot. Figure~\ref{fig:RobotHardware} shows the components of the robot. Six rechargeable batteries are used for power supply. An USB communication unit is used for programming and monitoring the control board. Two light sensors provide sensory input to the control board which controls the two motors and wheels through the motor driver component. Since the robot was designed to be used in different experiments~\cite{robotClass}, it can actually be equipped with many more sensors and therefore the controller board is more powerful than what would be required for the exercise presented here.

Nowadays materials to build these robots are abundant. For example, ready-to-use chassis can be acquired from online retailers such as Maker SHED (\url{http://www.makershed.com}) or Dwengo (\url{http://www.dwengo.com}). Tutorials on how to build robots are easily accessible as in Make magazine (\url{http://makezine.com}) or any of the many blogs on robotics. The approximate material cost for the robot presented here is EUR 140. Information about how to rebuild the robot and the required software libraries is available on Dorit Assaf's website (\url{http://www.embed-it.ch}).

\section{Robot software}

The Arduino project provides open source programming libraries and software development kits. Alternatively, the MATLAB language offers the ArduinoIO\footnote{MATLAB is a widespread scientific computing language, almost a standard in the scientific research community nowadays, \url{http://www.mathworks.com}.}, an easy to use programming interface. Below we show a snippet of the C code used for a controller that produces Braitenberg's vehicle 3 behavior. Lest the unexperienced user find the source code daunting, the Arduino project offers very easy tutorials to get started.

The digital output that controlled the wheels had an 8-bit resolution (it can produce 256 different values), therefore the speed of the motor is given by a number between 0 and 255, being 127 the middle value or half-speed. The preamble of the code includes our custom libraries needed and initializes sensors and motors. Next a function to set up the robot is defined, it initializes the default robot speed (127 = half-speed) and forward direction. After this function is executed, the continuous loop() routine starts. There, the sensor values of light sensor 1 and light sensor 2 are read and saved in the variables {\tt sensorValue1} and {\tt sensorValue2}. The sensor values range from 0 (dark) to 1023 (bright). The map() function, as its name indicates, maps the first two arguments (the sensor range [0,1023]), to the range [255,0]. This value will replace the default speed of the robot via the setSpeed() function, therefore, bright light will slow down the robot.

{\fontsize{9}{4}

\begin{Verbatim}[commandchars=\\\{\}]
\PY{c+c1}{// Include libraries with functions}
\PY{c+c1}{// for the specific sensors and motors}
\PY{k+cp}{#}\PY{k+cp}{include <LightSensor.h>}
\PY{k+cp}{#}\PY{k+cp}{include <DCMotor.h>}
\PY{c+c1}{// Define sensors and motors}
\PY{c+c1}{// Two sensors connected to pins 1 and 2.}
\PY{k+kt}{LightSensor} \PY{n}{lightSensor}\PY{p}{(}\PY{l+m+mi}{1}\PY{p}{,} \PY{l+m+mi}{2}\PY{p}{)}\PY{p}{;}
\PY{c+c1}{// Connect pins to motor driver component}
\PY{k+kt}{DCMotor} \PY{n}{motor1}\PY{p}{(}\PY{l+m+mi}{12}\PY{p}{,} \PY{l+m+mi}{8}\PY{p}{,} \PY{l+m+mi}{10}\PY{p}{)}\PY{p}{;}
\PY{k+kt}{DCMotor} \PY{n}{motor2}\PY{p}{(}\PY{l+m+mi}{18}\PY{p}{,} \PY{l+m+mi}{19}\PY{p}{,} \PY{l+m+mi}{11}\PY{p}{)}\PY{p}{;}
\PY{k+kt}{void} \PY{n+nf}{setup}\PY{p}{(}\PY{p}{)}
\PY{p}{\PYZob{}}
\PY{c+c1}{// This function is loaded} 
\PY{c+c1}{// at startup and after each reset}

  \PY{c+c1}{// Set default speed of the motors}
  \PY{n}{motor1}\PY{p}{.}\PY{n}{setSpeed}\PY{p}{(}\PY{l+m+mi}{127}\PY{p}{)}\PY{p}{;}
  \PY{n}{motor2}\PY{p}{.}\PY{n}{setSpeed}\PY{p}{(}\PY{l+m+mi}{127}\PY{p}{)}\PY{p}{;}
  \PY{c+c1}{// Set default direction of rotation}
  \PY{n}{motor1}\PY{p}{.}\PY{n}{setDirection}\PY{p}{(}\PY{n}{FORWARD}\PY{p}{)}\PY{p}{;}		
  \PY{n}{motor2}\PY{p}{.}\PY{n}{setDirection}\PY{p}{(}\PY{n}{FORWARD}\PY{p}{)}\PY{p}{;}
\PY{p}{\PYZcb{}}
\PY{k+kt}{void} \PY{n+nf}{loop}\PY{p}{(}\PY{p}{)}
\PY{p}{\PYZob{}}
\PY{c+c1}{// This function runs while the robot is alive}

  \PY{c+c1}{// Read sensor values}
  \PY{k+kt}{int} \PY{n}{sensorValue1} \PY{o}{=} 
  \PY{n}{lightSensor}\PY{p}{.}\PY{n}{readSensorValue1}\PY{p}{(}\PY{p}{)}\PY{p}{;}	
  \PY{k+kt}{int} \PY{n}{sensorValue2} \PY{o}{=} 
  \PY{n}{lightSensor}\PY{p}{.}\PY{n}{readSensorValue2}\PY{p}{(}\PY{p}{)}\PY{p}{;}
  
  \PY{c+c1}{// Convert sensor values to motor speed}
  \PY{k+kt}{int} \PY{n}{newSpeed1} \PY{o}{=} 
  \PY{n}{map}\PY{p}{(}\PY{n}{sensorValue1}\PY{p}{,} \PY{l+m+mi}{0}\PY{p}{,} \PY{l+m+mi}{1023}\PY{p}{,} \PY{l+m+mi}{255}\PY{p}{,} \PY{l+m+mi}{0}\PY{p}{)}\PY{p}{;}	
  \PY{k+kt}{int} \PY{n}{newSpeed2} \PY{o}{=} 
  \PY{n}{map}\PY{p}{(}\PY{n}{sensorValue2}\PY{p}{,} \PY{l+m+mi}{0}\PY{p}{,} \PY{l+m+mi}{1023}\PY{p}{,} \PY{l+m+mi}{255}\PY{p}{,} \PY{l+m+mi}{0}\PY{p}{)}\PY{p}{;}
  \PY{c+c1}{// Apply new speed values to motors}
  \PY{n}{motor1}\PY{p}{.}\PY{n}{setSpeed}\PY{p}{(}\PY{n}{newSpeed1}\PY{p}{)}\PY{p}{;}
  \PY{n}{motor2}\PY{p}{.}\PY{n}{setSpeed}\PY{p}{(}\PY{n}{newSpeed2}\PY{p}{)}\PY{p}{;}
\PY{p}{\PYZcb{}}
\end{Verbatim}

}

Litle more is needed to get the robot running. The source code is available on the website \url{http://www.embed-it.ch} together with some programming guidelines.

\section{Discussion and conclusion}

During the class we observed that students were fully engaged and were having fun. Based on their feedback, we attribute this to the presence of the robot, a non-standard tool for teaching. 

The students produced creative models (with a tendency to complicated schemes), hypotheses, and interesting experiments. No group actually found the correct solution (Braitenberg's vehicle 3) or anything equivalent. However, one group proposed a feedback controller that, despite its complexity, seemed aligned with Braitenberg's ideas. Nevertheless, our goal was to challenge the students and allow them to build their hypothesis based on hands-on evidence, therefore this goal was met. Using robots as a learning tool allows to prepare fully customizable class activities with different levels of difficulty and which can emulate real research situations.

The validation process, given that expectations were clearly stated, resulted to be fairly simple: either the model predicted the behavior or not. Several students showed determination to find a working model and automatically reworked theirs without being told to do so. We were surprised to note that the cycle: build a model, test it, rework the model; emerged naturally after the little push given when we described the activity to the students.

From the teacher's perspective this activity requires some extra work especially in the preparation phase. However, the effort was worth it and we encourage other teachers to try. A caveat of this kind of exercise is the difficulty to define criteria to grade a student's performance, due to the unstructured nature of the activity and the variety of possible solutions. This could be avoided by complementing the activity with a written report or a presentation. In the case where parallel activities are also performed such as calibration of sensors or construction of a speedometer\footnote{Students could build wheel speed sensors (speedometers) and verify the discussion in~\cite{SpeedM}}, grading could be simplified.

A more physics based experience could be to set up {\it a robotic car crash} and engage students in a forensic physics experience, where they could determine initial speeds and directions or maneuvers made by the {\it artificial car drivers}~\cite{ForFis}.

We invite other teachers to try similar activities. We offer our support for the programming and assembly of the robot and invite the reader send us feedback.


\end{document}